\documentclass[prl,twocolumn,showpacs,preprintnumbers,amsmath,amssymb,superscriptaddress,nofootinbib]{revtex4}
\usepackage{graphics}
\usepackage{graphicx}
\usepackage{epsfig}
\usepackage{epsf}
\usepackage{bm}
\usepackage{subfigure}
\usepackage{amsmath}
\usepackage{color}
\usepackage{float}
\usepackage{url}
\usepackage{hyperref}
\usepackage[utf8]{inputenc}
\usepackage{amssymb}

\usepackage{fancyhdr}
\usepackage{alltt}
\usepackage{comment}
\usepackage{url}

\newcommand{\bea}{\begin{eqnarray}}
\newcommand{\eea}{\end{eqnarray}}
\newcommand{\be}{\begin{equation}}
\newcommand{\ee}{\end{equation}}

\begin{document}

\title{Improving inverse Compton sources by removing non-linearities}



\author{B. Terzi{\'c}}
\affiliation{Department of Physics, Center for Accelerator Science, Old Dominion University, Norfolk, Virginia 23529, USA}

\author{G. A. Krafft}
\affiliation{Thomas Jefferson National Accelerator Facility, Newport News, Virginia 23606, USA}
\affiliation{Department of Physics, Center for Accelerator Science, Old Dominion University, Norfolk, Virginia 23529, USA}

\begin{abstract}
We present a new, more nuanced understanding of non-linear effects in inverse 
Compton sources. Contrary to what has been heretofore understood, deleterious
non-linear effects can arise even at low laser intensities, a regime previously viewed
as linear. After laying out a comprehensive survey of all non-linear phenomena which 
degrade the effectiveness of inverse Compton sources, we discuss two powerful 
techniques designed to remove these non-linearities. Starting with the known technique 
of longitudinal chirping of the laser pulse, which we developed 
earlier to mitigate adverse non-linear effects in the high laser field regime,
we have discovered that the simple stretching of the laser pulse, while keeping the
energy constant, can significantly increase the spectral density of the scattered radiation
in many operating regimes. 
Our numerical simulations show that combining these two techniques removes
detrimental non-linearities and improves the performance of inverse Compton sources
over an order of magnitude.
\end{abstract}


\maketitle

X-rays enable scientists to see the internal structure of materials on all length scales from the 
macroscopic down to the positions of individual atoms. This capability has had profound impact 
on science, technology, and on the world economy. It is impossible to overstate this impact, 
from Nobel Prize winning science to the everyday dental x-ray. The science and technology 
community agrees that future advances in many areas depend on understanding 
structure/function relationships at the nano-scale where new properties emerge, and controlling 
the fabrication of complex materials at that scale to achieve transformative physical, chemical, 
and biological functionality.

The sources of x-ray radiation relying upon Compton scattering \cite{jackson,priebe} possess 
a notable advantage over the traditional bremsstrahlung sources--the narrow-band nature of 
the radiation emerging from them. This motivated creation of designated facilities featuring
inverse Compton sources (ICS) \cite{huang}, which have been applied to x-ray structure 
determination \cite{aetal2010}, dark-field imaging \cite{betal2009,setal2012}, phase contrast 
imaging \cite{betal2009}, and computed tomography \cite{aetal2013} .

As a significantly more affordable alternative to large facility sources, 
inverse Compton sources (ICS) of x-rays may allow a multitude of studies not easily accomplished 
at the large facilities. These potentially groundbreaking studies in medicine, pharmaceutical 
industry, chemistry, material science, homeland security and many other fields of human 
endeavor, hold a promise of fundamentally improving our lives. However, in order to bring 
about this new era, we first need to develop a deeper understanding of physical processes 
in ICS.

In this letter, we map out the regimes of operations for ICS and describe the non-linear effects 
which plague each of these regimes. We then present two precise techniques which combine 
to largely remove these non-linearities, thereby substantially improving the peak spectral 
density of the scattered radiation emerging from these sources.
\begin{figure}[htb]
\includegraphics[width=3.3in,height=2.35in]{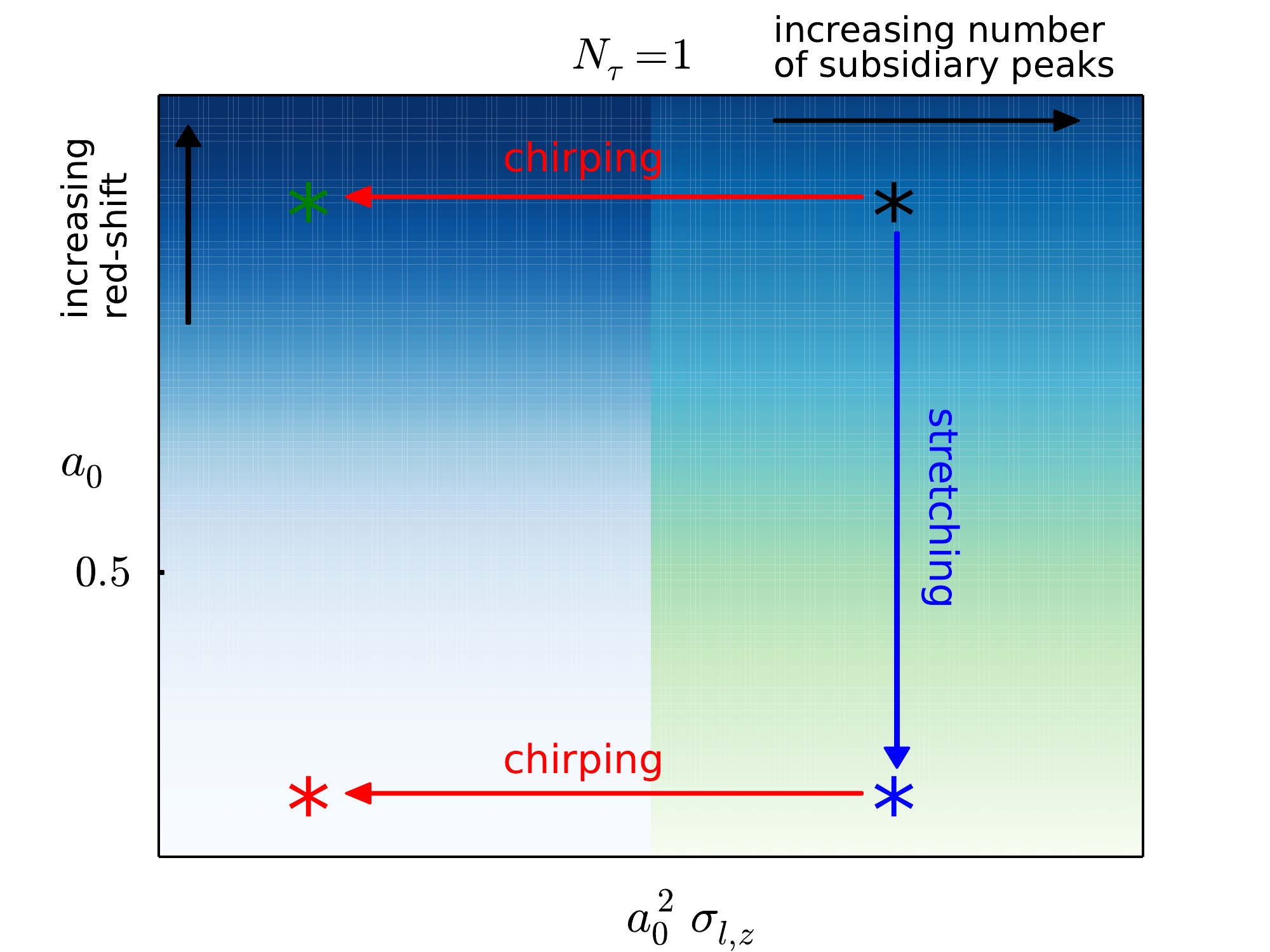}
\caption{\small{Regimes of operation of ICS, as listed in Table \ref{tab0}. Blue line pointing
downward represents stretching the laser pulse (increasing its length $\sigma_{l,z} $), while 
keeping the energy of the laser constant ($a_0^2 \sigma_{l,z} = const.$). Red lines pointing to 
the left represent chirping the laser pulse, which removes subsidiary peaks. Stars denote 
parameters for simulations carried out in Fig.~\ref{fig_spectra1} and \ref{fig_spectra2}, each color 
corresponding to the spectrum of the same color.}}
\label{fig1}
\end{figure}

In ICS, a relativistic electron beam interacts with a high-field laser beam, producing intense and 
highly collimated electromagnetic radiation via Compton scattering \cite{priebe}. Through 
relativistic upshifting and the relativistic Doppler effect, highly energetic polarized photons are 
radiated along the electron beam motion when the electrons interact with the laser light.

The strength of the laser field is quantified by the amplitude ($a_0$) of the normalized vector 
potential ${\tilde A}(\xi) = e A (\xi)/m_e c = a(\xi)\cos(2\pi \xi/\lambda)$, where $a(\xi)$ is the 
amplitude envelope, $\xi = z + ct$ is the coordinate along the laser pulse, $\lambda$ the wavelength 
of the laser. At high laser field intensities, the spectrum of backscattered radiation is considerably 
broadened because of the changes in the longitudinal velocity of the electrons during the pulse.
Such ponderomotive line broadening in the scattered radiation makes the bandwidth too large 
for some applications and reduces the spectral brilliance.

The onset of non-linear ponderomotive line broadening has been associated with the increase
of the laser field strength \cite{k2004,tdhk2014,tetal2019}. However, we present a more nuanced
analysis of the origins of non-linearities which fall under the umbrella of
ponderomotive broadening. This analysis, illustrated in Fig.~\ref{fig1} and detailed in the text,
helps us devise ways of removing non-linearities from ICS.

The two main mechanisms which bring about non-linear ponderomotive broadening of the 
spectra are: (1) non-linear subsidiary peaks and (2) non-linear redshift. The interplay between
the two effects is shown in Fig.~\ref{fig1}, with each axis quantifying the importance of each.

Linear spectrum features solitary, narrow peaks. One of the signatures of spectral non-linearities 
is the appearance of multiple peaks around each harmonic. These non-linear subsidiary peaks in 
the scattered spectra at high laser intensities been observed and their number empirically 
approximated for a gaussian laser envelope and $\lambda=800$ nm in 
Refs.~\cite{hsk2010,gsu2013}. The exact number of subsidiary peaks for an arbitrary laser envelope
and wavelength was derived in Ref.~\cite{tdhk2014}. For a gaussian envelope we consider here,
it is
\begin{equation} \label{Ntau}
N_{\tau} = {{\sqrt{\pi}}\over{2}} a_0^2 {{\sigma_{l,z}}\over{\lambda}},
\end{equation}
where $\sigma_{l,z}$ is the length of the laser pulse. For $N_{\tau} > 1$, 
additional peaks appear in the spectrum, thereby reducing the spectral density in the primary peak.

\begin{figure*}[htb]
\includegraphics[width=2.32in]{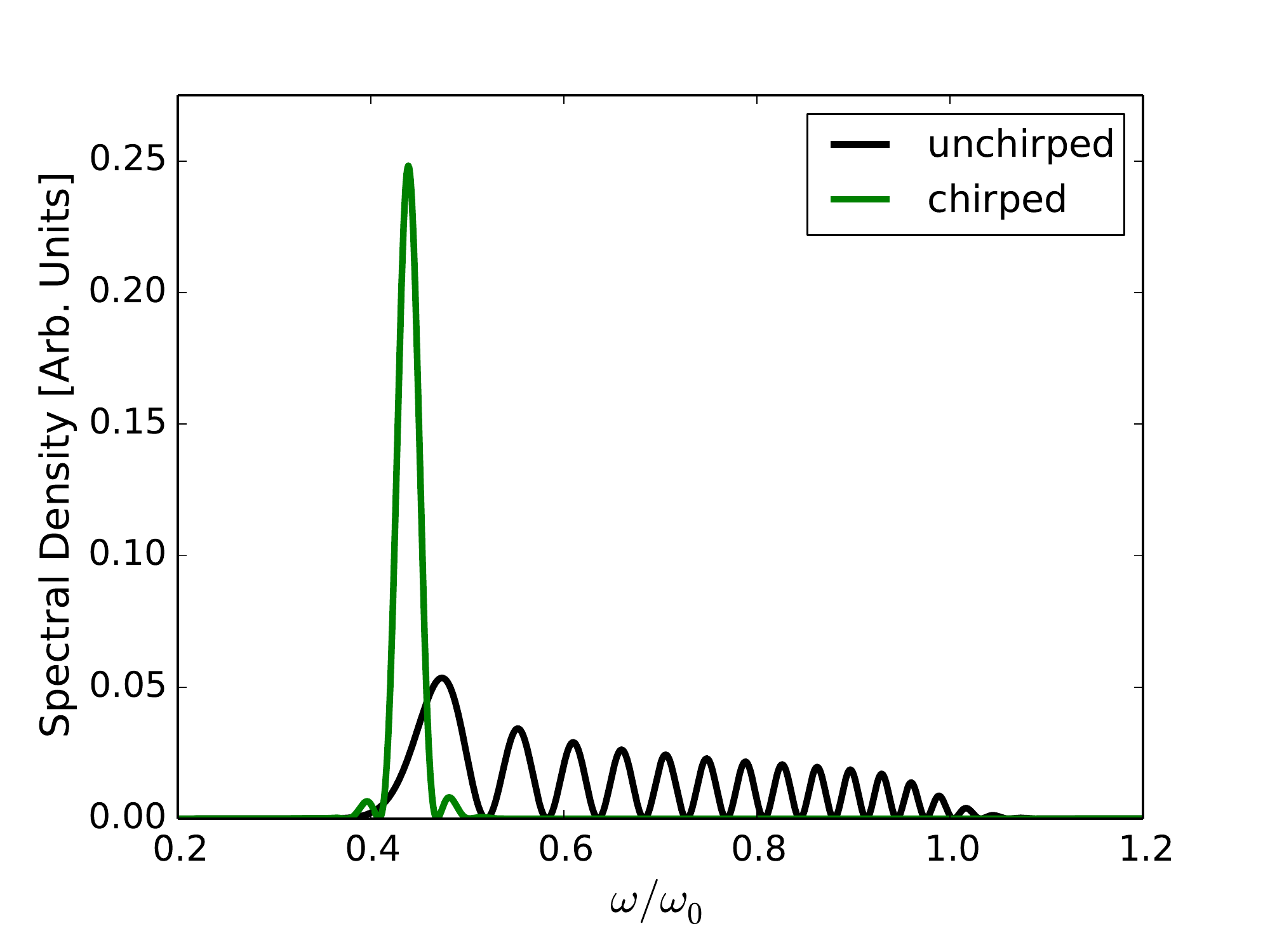}
\includegraphics[width=2.32in]{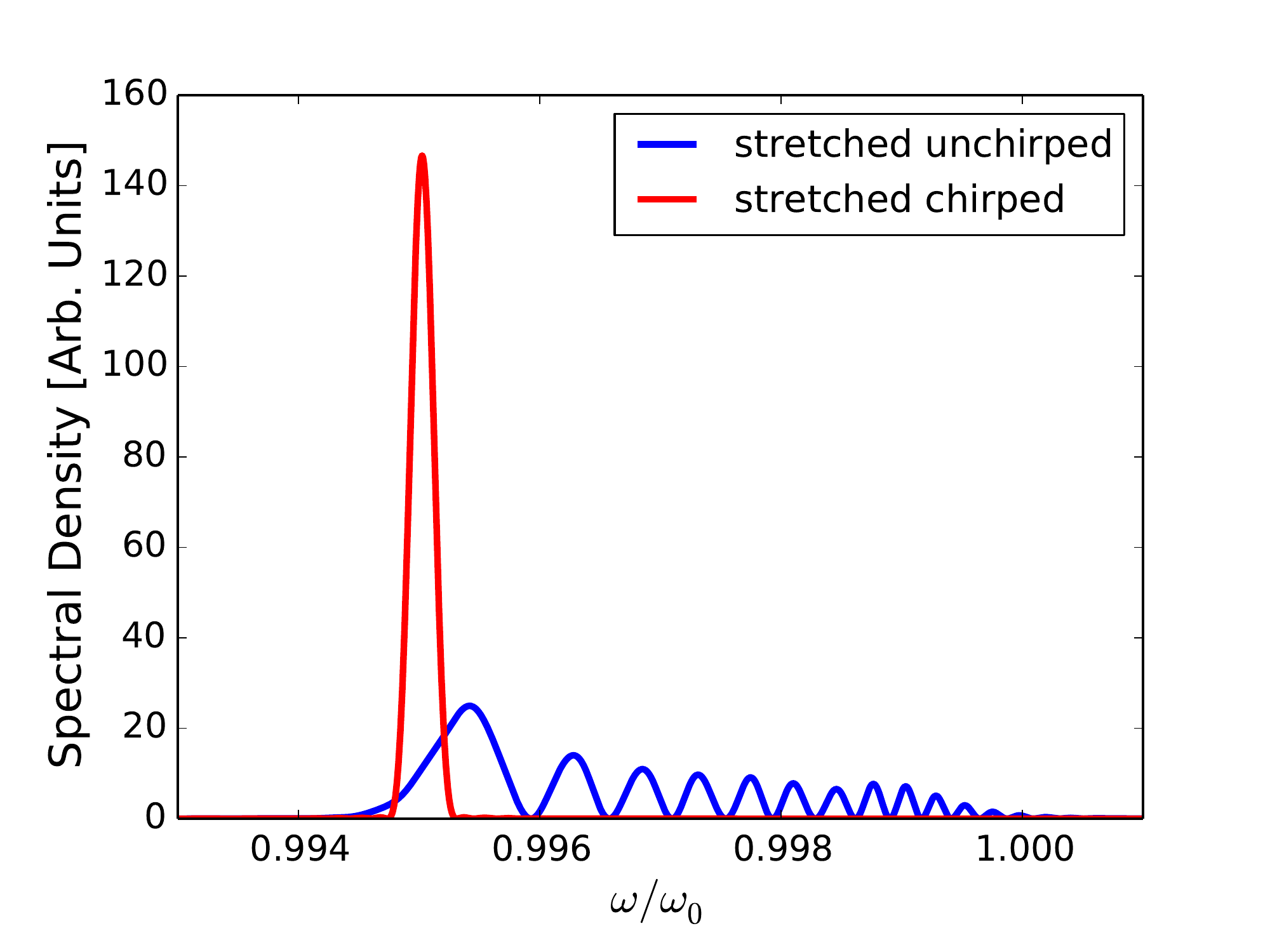}
\includegraphics[width=2.32in]{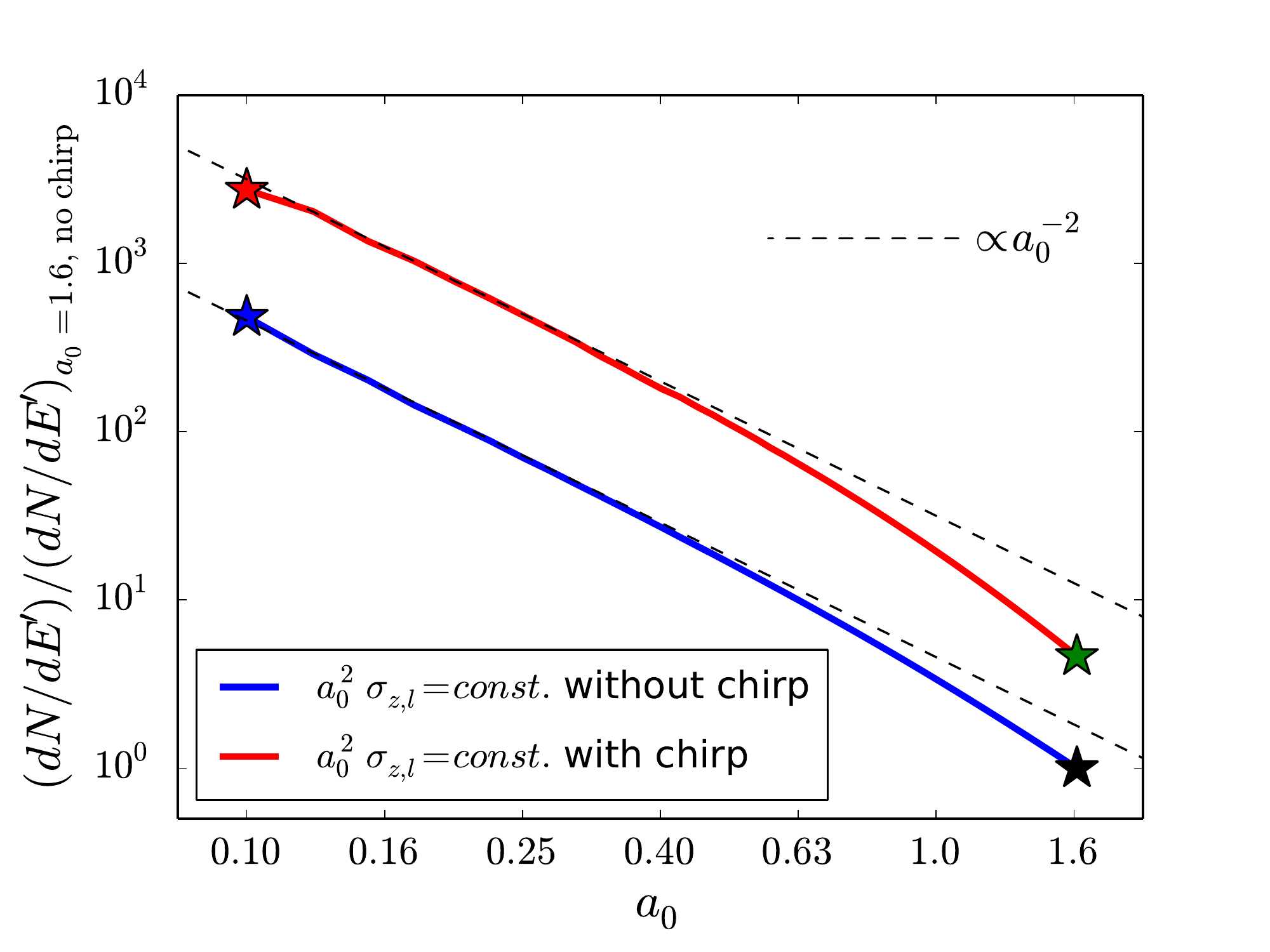}
\caption{\small{Backscattered spectra for a single, on-axis 45 MeV electron colliding 
with a 1D plane wave laser with wavelength of $\lambda=800$ nm. 
Panels (a) and (b) use the same arbitrary units.
(a): Unchirped (black line) and chirped (green line) laser with $a_0 = 1.6$, 
$\sigma_{l,z} = 4.46~\mu$m. Chirping increases the peak spectral density by 4.6 times.
(b): Unchirped (blue line) and chirped (red line) laser with $a_0 = 0.1$, 
$\sigma_{l,z} = 1.14~$mm.
Chirping increases the peak spectral density by 5.3 times.
Stretching increases the peak spectral density by 480 times
(unchirped laser) and 591 times (chirped laser). Combined,
stretching and chirping increase the spectral density by 2738 times.
(c): The relative increase in peak spectral density,
normalized to the spectral density of the unchirped case at $a_0=1.6$,
as a function of the laser field strength parameter $a_0$.
}}
\label{fig_spectra1}
\end{figure*}

From the Eq.~(\ref{Ntau}), it follows that the number of subsidiary peaks does not depend on 
the strength of the laser field $a_0$ alone, but on the combination of the strength of the laser field 
and the length of the laser, $a_0^2 \sigma_{l,z}$. Therefore, it is possible for a spectrum to
feature non-linear subsidiary peaks even at low $a_0$, a regime previously viewed as linear, if the 
laser is long enough. The spectrum shown in blue in Fig.~\ref{fig_spectra1}b, which 
represents $a_0=0.1$, is an illustration of such a case.

Removing subsidiary peaks and restoring the narrow-band nature of the spectrum is accomplished
by exact laser chirping, which changes the normalized vector potential to
${\tilde A}(\xi) =  a(\xi)\cos(2\pi \xi f(\xi) /\lambda)$, where $f(\xi)$ is the chirping prescription
\cite{tdhk2014}:
\begin{equation} \label{f_exact}
f (\xi; a_0) = {{1}\over{1+a_0^2/2}} \left(1 + {{1}\over{2\xi}} \int_0^\xi a^2(\xi') d\xi' \right).
\end{equation}
Exact laser chirping was shown to be perfectly effective 
for restoring narrow-band property of the scattered radiation produces in scattering a 1D plane
wave laser with an on-axis, single-electron \cite{tdhk2014} and an electron beam with an energy
spread \cite{trk2016}. 
Improving the laser model to a 3D laser pulse, brought about a generalized chirping 
prescription when colliding with a single on-axis electron \cite{metal2018}, which, while 
not perfect, is still quite effective. 
Finally, in a realistic scenario of a collision between a 3D laser pulse and a general electron 
beam distribution, we have shown that the optimal form of laser chirping can be found using 
an optimization over the parameters of the chirping prescription \cite{tetal2019}. 
The effectiveness of such chirping depends on the relative transverse sizes of the two 
colliding beams and can still be quite effective when the transverse size of the laser pulse is  
roughly the size of the electron beam or larger \cite{tetal2019}.

Therefore, the laser chirping removes subsidiary peaks, leaving only one dominant peak per
harmonic. The effect on laser chirping on the landscape of regimes of operation for ICS 
shown in Fig.~\ref{fig1} is to move horizontally from the region where $N_{\tau} \gg 1$ to
the region where $N_{\tau} \approx 1$.

Linear spectrum contains a dominant, narrow peak centered around the frequency
$\omega_0 = (1+\beta)^2 \gamma^2 2 \pi c/\lambda$. As the magnitude of the 
normalized vector potential $a_0$ grows, the peak experiences non-linear red-shift:
\begin{equation} \label{shift}
\omega = {{\omega_0}\over{1+ a_0^2/2}}.
\end{equation}
The non-linear shift is accompanied by widening of the peaks. This is evident from the 
comparison of the location of the primary harmonic in spectra in 
Figs.\ref{fig_spectra1}a-\ref{fig_spectra1}b.

From the Eq.~(\ref{shift}), it is clear that the non-linear shift will be reduced if $a_0$ is reduced.
If the total energy in the laser pulse is kept constant, the reduction in $a_0$ can only 
come at an appropriate increase in the laser volume $V$, because the laser intensity
$I$ is given by 
\begin{equation}
I = {{c E}\over{V}} = {{2 \pi^2 m_e^2 c^3}\over{\lambda^2 e^2 \mu_0}} a_0^2,
\end{equation}
where $E$ is the laser energy
\begin{equation}
E = {{2 \pi^2 m_e^2 c^2}\over{\lambda^2 e^2 \mu_0}} a_0^2 V.
\end{equation}
For a gaussian laser pulse, both transversally and longitudinally, 
$V = (2\pi)^{3/2} \sigma_{l,x} \sigma_{l,x} \sigma_{l,z}$, where $\sigma_{l,x}$ is the horizontal
and $\sigma_{l,y}$ vertical size of the laser pulse. $m_e$ is the mass of the electron, 
$c$ is the speed of light $e$ is the elementary charge and $\mu_0$ is the permeability of vacuum.
Therefore, to keep energy constant, reducing $a_0$ must be done in such a manner that the quantity
$a_0^2 V$ is kept constant. This means that reducing $a_0$ by some factor is accomplished by making
the volume (the transverse or longitudinal size, or combination thereof) larger by a square of that factor. 
In the case of 1D laser pulse model, the only way to increase the volume of the laser is by 
stretching it (increasing its longitudinal size). In the case of a 3D laser pulse model, the increase 
in volume can be done by either stretching the beam or increasing its cross-section 
(transverse size). However, increasing the cross-section of the laser while keeping the
energy constant will result in the net reduction of peak spectral density. While increasing the 
transverse size of the laser pulse leads to a more favorable distribution of effective laser field 
strength parameter $a_0$ that electrons experience \cite{tetal2019}, the peak spectral density 
will be dominated by its dependence on $a_0^2 \sigma_{l,z}^2$ \cite{trk2016}. 

This realization suggests a simple method of reducing the deleterious effects of non-linear red-shift: 
stretching the laser beam while keeping its energy constant. The effects of this approach on
peak spectral density are presented next, first for the 1D plane wave and then for the 3D pulse 
laser model.

\begin{table}
\begin{center}
\footnotesize
        \setlength\tabcolsep{4pt}
        \begin{tabular}{|l |c |c |l |}
                \hline 
                {\bf Regime} & {\bf $a_0^2 s$} & {\bf $a_0$} & {\bf Non-linearities} \\
                \hline
                Linear (white) & $N_\tau \lesssim 1$ & $a_0 \lesssim 0.5$ & none \\
                Non-linear (green) & $N_\tau \gtrsim 1$ & $a_0 \lesssim 0.5$ & subs.~peaks  \\
                Non-linear (blue) & $N_\tau \lesssim 1$ & $a_0 \gtrsim 0.5$ & red-shift \\
                Non-linear (aqua)& $N_\tau \gtrsim 1$ & $a_0 \gtrsim 0.5$ & red-shift, subs.~peaks \\
                \hline
        \end{tabular}
        \caption{Regimes of operations for ICS, as shown in Fig.~\ref{fig1}, color-coded.}
        \vskip-10pt
\label{tab0}
\normalsize
\end{center}
\end{table}

We verify these ideas through numerical simulations. The theory and the code for an on-axis, 
single-electron collision with a 1D laser pulse was developed in Ref.~\cite{k2004}. 
In Ref.~\cite{tdhk2014}, we modified it to model an arbitrary laser chirp. The code has been
further improved to carry out simulations for electron beams with energy spread \cite{trk2016}.
Finally, we recently developed another code, still based on the original theory developed for a 
single-electron scattering off a 1D laser pulse \cite{k2004,tdhk2014}, to model scattering of a
3D laser pulse (chirped or unchirped) off a realistic electron beam (with emittances and energy 
spread). This is the most general code, named SENSE (Simulation of Emitted Non-linear 
Scattering Events), subsuming all previous versions. It is described in detail in Ref.~\cite{tetal2019}. 
It is valid in a regime where electron recoil can be neglected, or, in terms of the subsidiary
peaks, $N_\tau \le 3/(4\alpha) \approx 103$, where $\alpha$ is the fine structure constant. This is not
a very restrictive range, because most, if not all, existing and future ICS are situated 
comfortably within this regime.
SENSE can be used to compute a spectrum from a single, on-axis electron scattering off a 1D 
plane wave (using a 3D model in the 1D limit $\sigma_{l,x}, \sigma_{l,y} \gg \sigma_{l,z}$). 
We can also use it for fully realistic simulations---including electron beams with energy 
spread and emittance and a 3D laser pulse with finite transverse extent---of an existing ICS. 

The individual effects of stretching of the laser and chirping are best visualized separately.
We do that by following the path marked by the colored stars in Fig.~\ref{fig1}. We start with an
unchirped case with high $a_0$ (the black star in Fig.~\ref{fig1} and the black line in 
Fig.~\ref{fig_spectra1}a) and stretch the laser pulse to the unchirped case 
with a low $a_0$ (the blue star in Fig.~\ref{fig1} and the blue line in 
Fig.~\ref{fig_spectra1}b), shown in Fig.~\ref{fig_spectra1}a. 
We then chirp the laser for both the low-$a_0$ (the red star Fig.~\ref{fig1} and the red 
line in of Fig.~\ref{fig_spectra1}b), and the high-$a_0$ case 
(the green star Fig.~\ref{fig1} and the green line in Fig.~\ref{fig_spectra1}a), 
shown in Fig.~\ref{fig_spectra1}b.

While at constant energy ($a_0^2 \sigma_{l,z} = const.$; moving vertically in Fig.~\ref{fig1}), the 
spectra look similar in shape when scaled by the non-linear offset $1/{(1+ a_0^2/2)}$. In the
low-$a_0$ regime, the height of the peak is directly proportional to 
$a_0^2 \sigma_{l,z}^2$ \cite{trk2016}:
\begin{equation}
{{d^2 E}\over{dE' d\Omega}} = {{\pi \alpha}\over{2}} \gamma^2 (1+\beta)^2 a_0^2 
{{\sigma_{l,z}^2}\over{\lambda^2}},
\end{equation}
where it is assumed that the collecting aperture is small enough that the spread it generates
is less then the spread from the pulse and the spectral height is less than the pedestal level
\cite{ketal2016}.
Therefore, at constant $a_0^2 \sigma_{l,z}$, the peak spectral density scales as $a_0^{-2}$. 
This dependence is verified in simulations for $a_0 \lesssim 0.5$, carried out for the fixed laser 
energy ($a_0^2 \sigma_{l,z} = const.$) and is shown in Fig.~\ref{fig_spectra1}c, 
for both unchirped and chirped lasers. For $a_0\gtrsim 0.5$, the drop in peak spectral density 
with increasing $a_0$ is even more precipitous because of (1) the relative importance of the 
higher-order harmonics increases with $a_0$, from being negligible at $a_0 \lesssim 0.5$ to 
rivaling the primary harmonic at $a_0 \gtrsim 1$ \cite{trk2016}; and (2) the complete merger 
of all harmonics at $a_0 \gtrsim 2$.

Taking a couple of steps forward in terms fidelity of computational modeling to the physical
problem, we carry out numerical simulations with SENSE for a collision of an electron 
beam---specified by its bulk properties of emittance, size and energy spread---and a 3D 
laser pulse. We use parameters from an actual experiment, and simulate what would happen 
if the laser in the experiment was changed--stretched and/or chirped. Chirping is implemented 
by varying $a_0$ in Eq.~(\ref{f_exact}).

We model an experiment at the ICS in Dresden, reported in Ref.~\cite{ketal2018},
which we recently simulated in detail \cite{tetal2019}. In our simulation, the electron beam 
is kept constant, possessing the same properties as the electron beam in the Dresden 
experiment \cite{ketal2018}.
Here we focus on the largest laser field strength reported, $a_0=1.6$, because in this case the 
non-linearities are most pronounced.

\begin{figure*}[htb]
\includegraphics[width=2.32in]{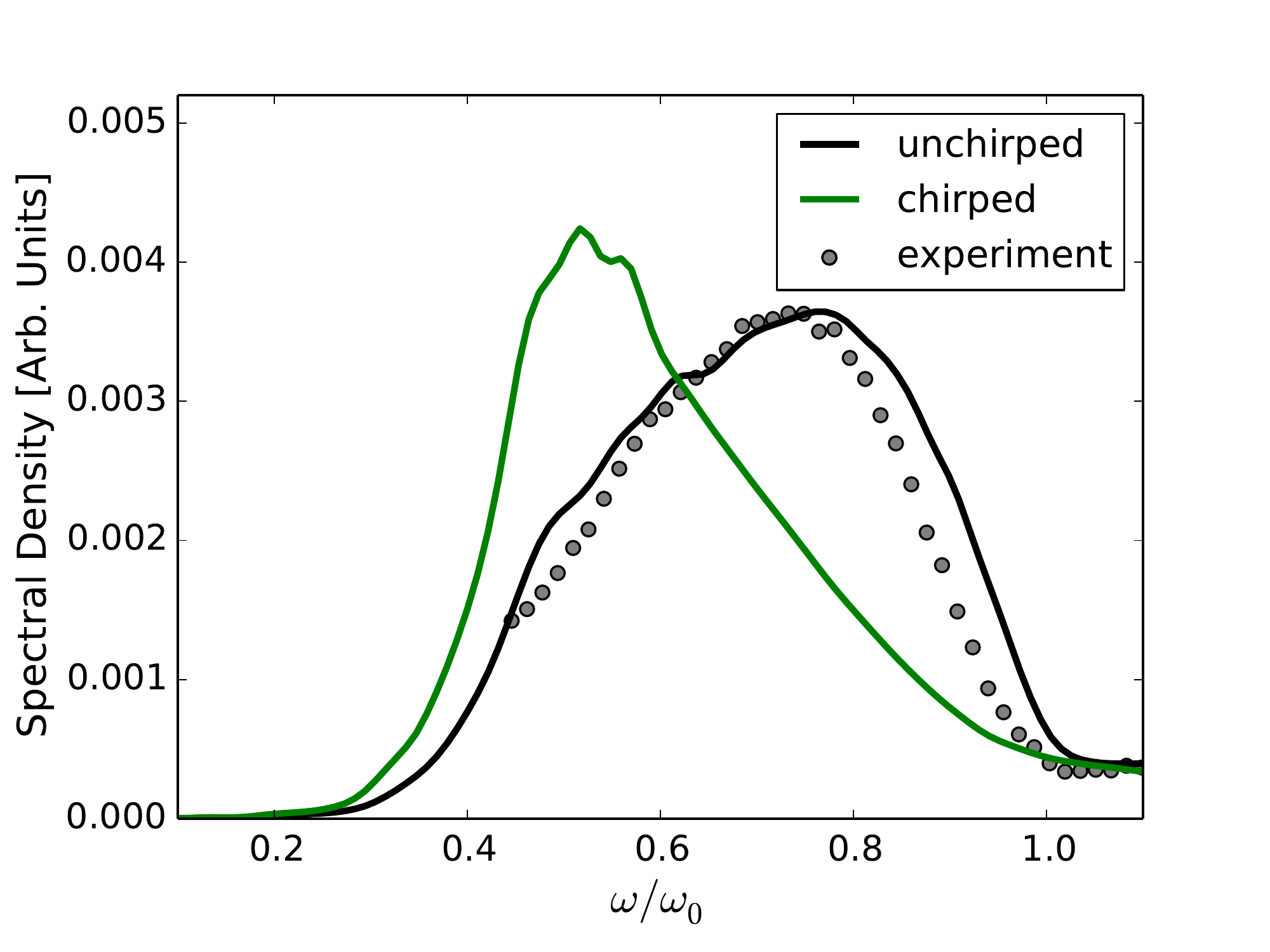}
\includegraphics[width=2.32in]{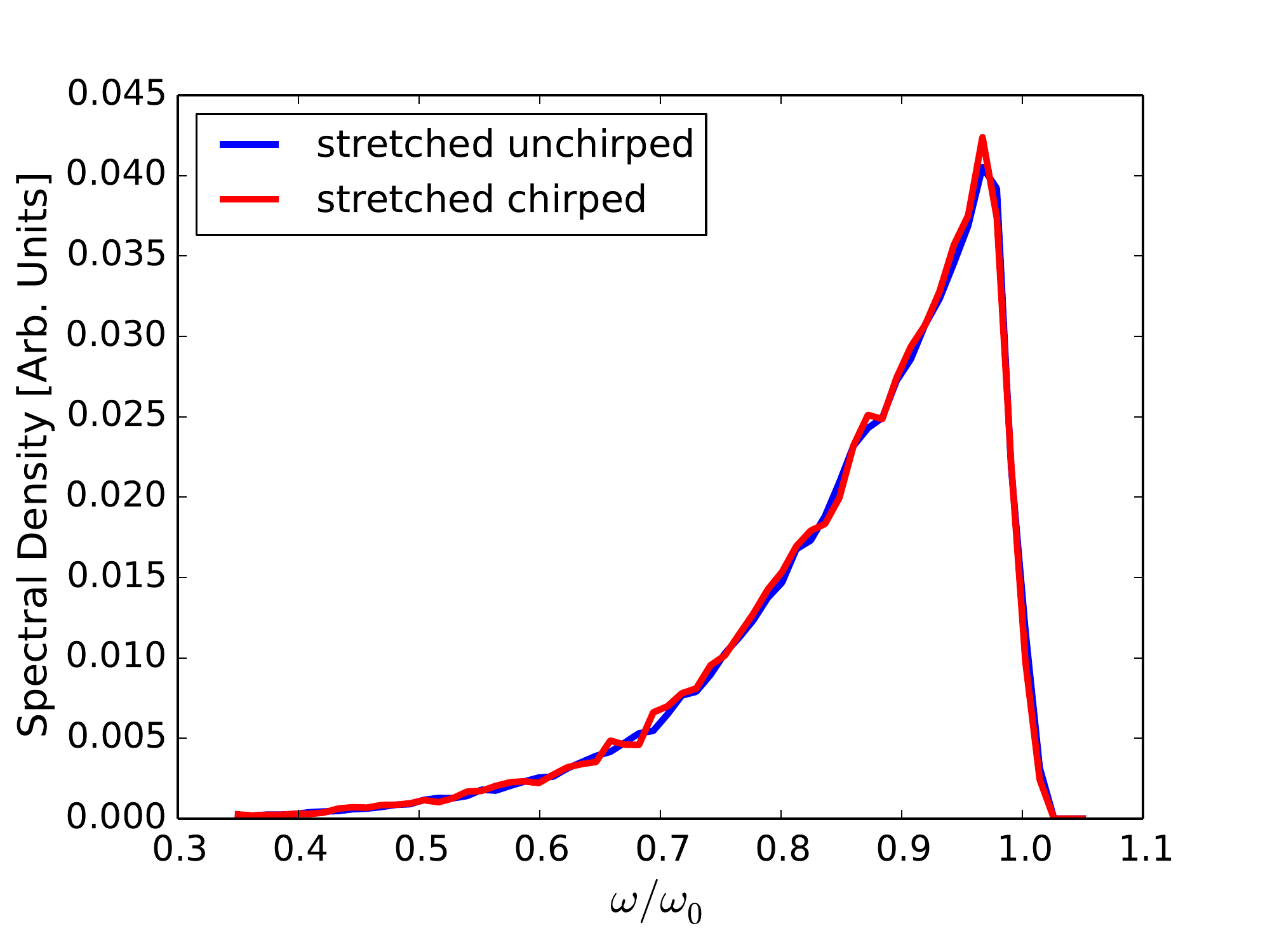}
\includegraphics[width=2.32in]{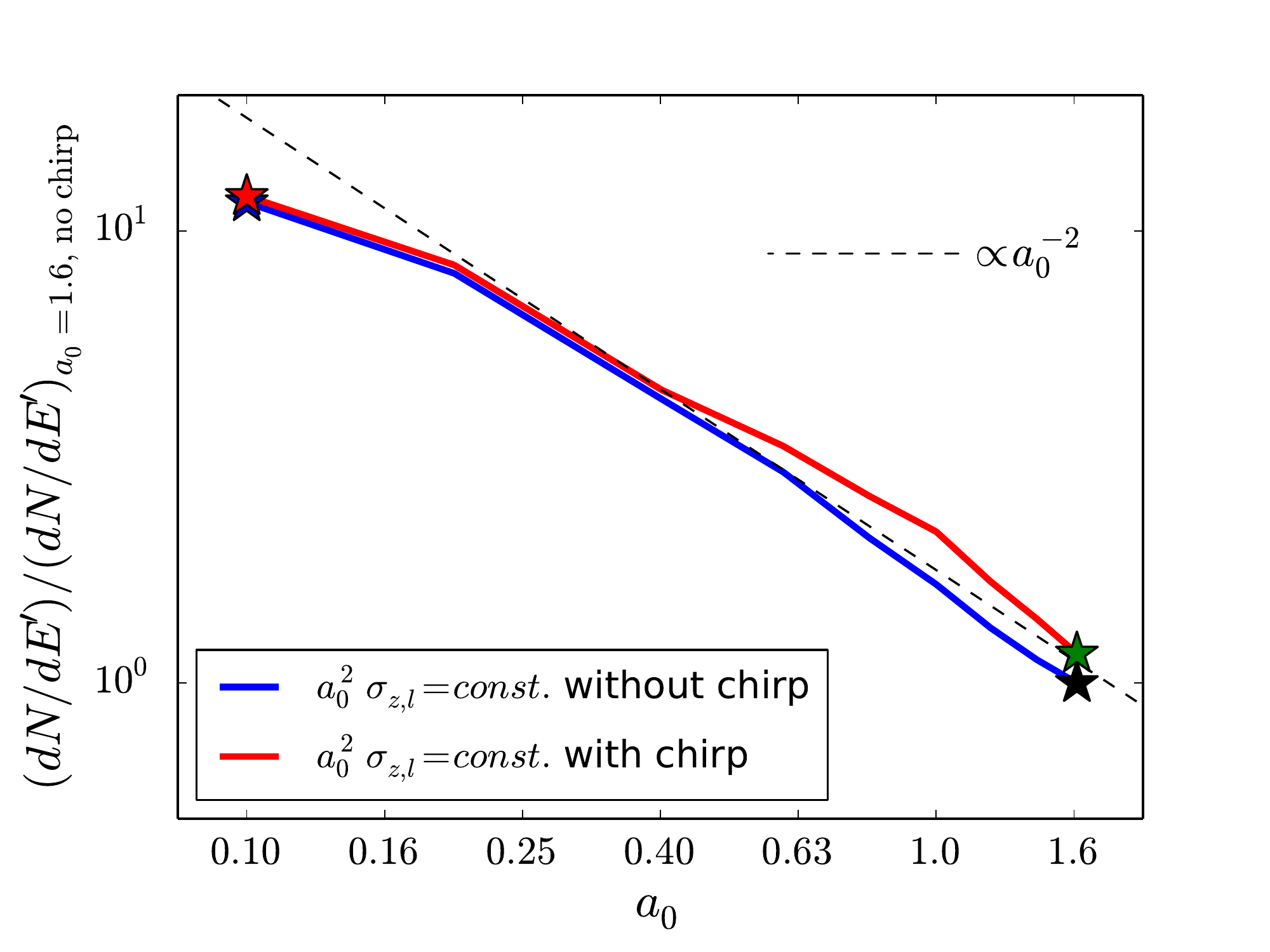}
\caption{\small{Spectra for a scattering event as for $a_0=1.6$ in Ref.~\cite{ketal2018}:
45 MeV electron beam with normalized transverse emittances $\epsilon_{x,n} = 20.3$ mm mrad, 
$\epsilon_{x,n} = 18$ mm mrad, transverse size $\sigma_x = 41~\mu$m, 
$\sigma_y = 81~\mu$m, and the energy spread of $0.00175$, colliding 
with a 3D plane wave laser with wavelength of $\lambda=800$ nm and transverse
size $\sigma_{l,x} = \sigma_{l,y} = 13.59~\mu$m.
Panels (a) and (b) use the same arbitrary units.
(a): Unchirped (black line) and chirped (green line) laser with $a_0 = 1.6$; 
experimental data from Ref.~\cite{ketal2018} (grey points). 
Chirping increases the peak spectral density by 16\%.
(b): The laser pulse is stretched longitudinally so as to keep $a_0^2 \sigma_{l,z}=const.$:
unchirped (blue line) and chirped (red line) laser with $a_0 = 0.1$ (blue line)
and $\sigma_{l,z} = 1.14~$mm.
Chirping is a non-factor.
Stretching increases the peak spectral density by about 12 times for both unchirped and chirped lasers.
(c): The relative increase in peak spectral density,
normalized to the spectral density of the unchirped case at $a_0=1.6$,
as a function of the laser field strength parameter $a_0$.
}}
\label{fig_spectra2}
\end{figure*}

Again, we visualize the individual effects of stretching of the laser and chirping separately
by following the path marked by the colored stars in Fig.~\ref{fig1}. 
We do that by following the path marked by the colored stars in Fig.~\ref{fig1}. We start with an
unchirped case with high $a_0$ (the black star in Fig.~\ref{fig1} and the black line in 
Fig.~\ref{fig_spectra2}a) and stretch the laser pulse to the unchirped case 
with a low $a_0$ (the blue star Fig.~\ref{fig1} and the blue line in  
Fig.~\ref{fig_spectra2}b), shown in Fig.~\ref{fig_spectra2}a. 
We then chirp the laser for both the low-$a_0$ (the red star Fig.~\ref{fig1} and the red 
line in Fig.~\ref{fig_spectra2}b), and the high-$a_0$ case 
(the green star Fig.~\ref{fig1} and the green line in Fig.~\ref{fig_spectra2}a), 
shown in Fig.~\ref{fig_spectra2}b.

The improvements in peak spectral density due to chirping are considerably more modest
for scattering off an electron beam than in the case of a single, on-axis electron. 
As the electrons from a beam distributions collide with a laser pulse, they will experience a 
wide range of laser field strength parameters $a_0$ \cite{tetal2019}. Exact chirping prescription 
in Eq.~(\ref{f_exact}) depends on $a_0$, and can perfectly compensate the non-linearities in 
the spectrum only for electrons experiencing one value of the laser field strength. 
Therefore, the majority of the electron distribution does not see the perfect chirping prescription. 

As the laser field strength $a_0$ reduces while keeping the energy constant, the longitudinal 
size of the laser $\sigma_{l,z}$ increases, and the energy spread of the pulse decreases. 
Stretching the laser pulse reduces its energy spread---in both chirped and unchirped 
case---and its contribution to the scattered linewidth becomes negligible when compared to 
the other sources, such as electron beam emittance. This is indeed observed from 
Fig.~\ref{fig_spectra2}b: a spectrum dominated by electron beam emittance where 
chirping is a non-factor. A comprehensive discussion of the scaling laws in ICS is presented in 
Refs.~\cite{setal2009,ketal2016,cetal2017,retal2018}, along with clear illustrations of what 
spectra dominated by various sources of linewidth look like. The asymmetry and the hard
high-energy edge in the spectrum in Fig.~\ref{fig_spectra2}b are clear
signs of dominanice of electron beam emmitance.

In Ref.~\cite{tetal2019}, we demonstrate that the effectiveness of chirping in improving the 
peak spectral density becomes more pronounced as the electron beam emittance is reduced.
This is to be expected, since reducing the beam emittance, reduces its size relative to that
of the laser pulse, thereby approaching the 1D plane wave limit \cite{tetal2019}.

The impact of laser stretching on peak spectral density is still substantial. In the case simulated
here, beam stretching from the nominal experimental setup for $a_0=1.6$ to $a_0=0.1$ 
increases the peak spectral density by about a factor of 10.

\begin{figure}[htb]
\includegraphics[width=3in]{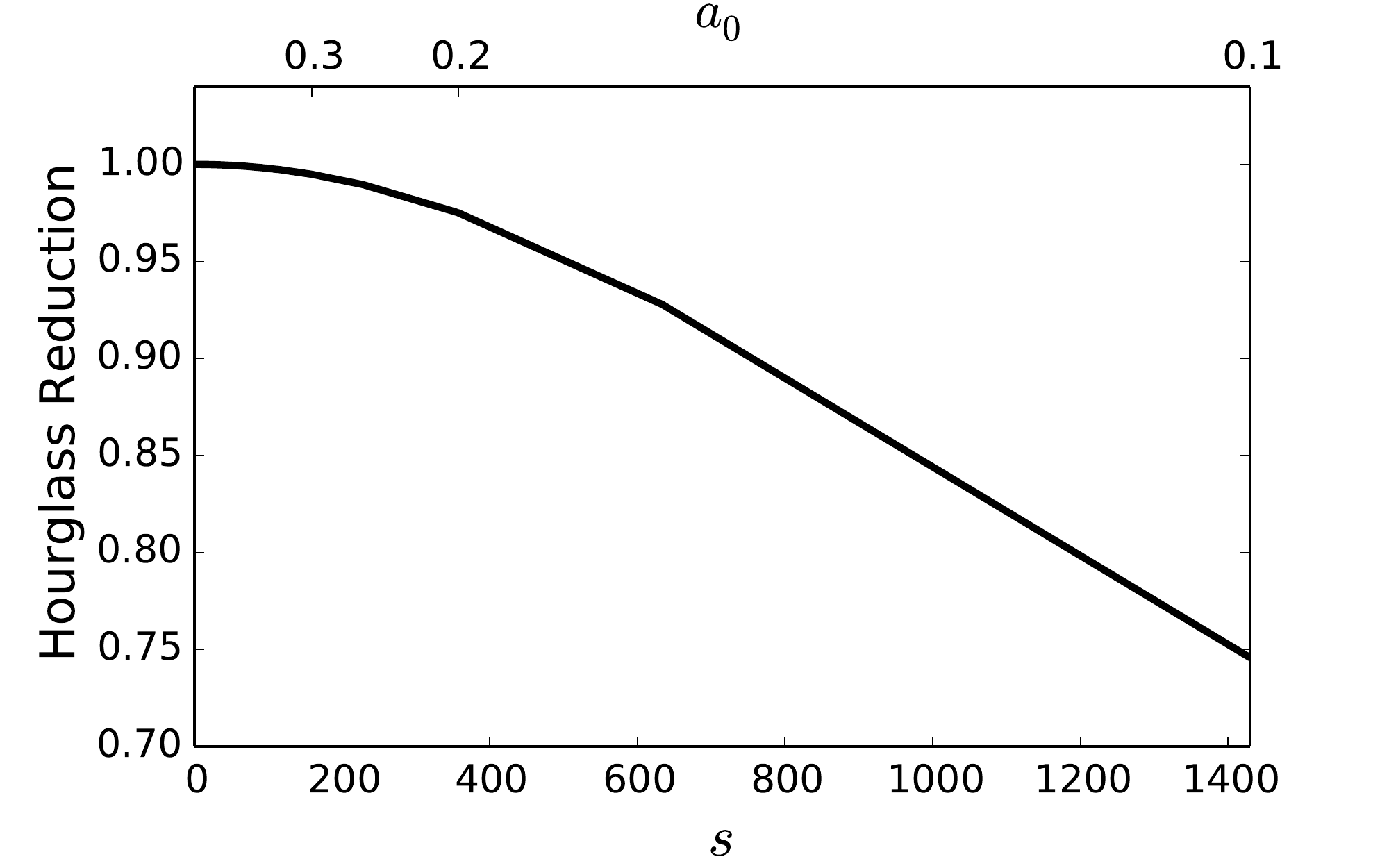}
\vskip-5pt
\caption{\small{Hourglass reduction factor as a function of the size of the laser pulse, normalized
to the value at nominal parameters $a_0=1.6$ and $s=5.57$.}}
\label{fig_hg}
\end{figure}

As the laser pulse is stretched, it experiences an hourglass effect which reduces
the rate of interaction \cite{f1991}. The reduction with respect to the nominal parameters
at $a_0=1.6$ and $\sigma_{l,z}=4.46~\mu$m is shown in Fig.~\ref{fig_hg}. For the case 
considered here, it is not significant---on the order of 30\% at $a_0=0.1$. 

We put forward a new, more subtle, understanding of non-linearities in ICS. 
One novel realization is that non-linear subsidiary peaks in the scattered spectra can 
arise even at low values of the laser field strength parameter $a_0$, a region previously 
thought of as linear.
Another new idea presented here, more profound and consequential in magnitude, is that the
deleterious non-linear redshift can be removed by stretching the laser pulse while
keeping its energy constant. This stretching alone improves the peak spectral density
by over an order of magnitude. In cases when the linewidth of scattered radiation is not
dominated by the electron beam emittance, laser chirping can be useful removing 
subsidiary peaks from individual spectra, thereby further improving the peak spectral
density. Overall, combining these two approaches---stretching and chirping of the laser 
beam---removes both non-linear manifestations which are detrimental to spectral
linewidth: non-linear red-shift and subsidiary peaks, thereby substantial improving the 
performance of ICS.

The most important implication of our findings is that the ICS do not need high-field lasers 
to produce high-yield radiation. In fact, given the same laser energy, operating at lower field 
strengths increases the peak yield by a substantial factor, often exceeding an order of magnitude.
IOur findings substantially reduce the challenge of designing and operating high-field lasers, 
thereby lowering their cost and increasing the prospect of their wide-spread availability.

This paper is authored by Jefferson Science Associates, LLC under U.S.~DOE Contract 
No.~DE-AC05-06OR23177. B.~T. acknowledges the support from the U.S.~National 
Science Foundation award No.~1535641 and No.~1847771.


\begin{thebibliography}{10}

\bibitem{jackson}
J.~D. Jackson.
\newblock {\em Classical Electrodynamics}.
\newblock Wiley, 2007.

\bibitem{priebe}
G.~A. Krafft and G.~Priebe.
\newblock {\em Reviews of Accelerator Science and Technology}, 03(01):147--163,
  2010.

\bibitem{huang}
Z.~Huang and R.~D. Ruth.
\newblock {\em Physical Review Letters}, 80:976--979, Feb 1998.

\bibitem{aetal2010}
J.~Abendroth, M.~S. McCormick, T.~E. Edwards, B.~Staker, R.~Loewen, M.~Gifford,
  J.~Rifkin, C.~Mayer, W.~Guo, Y.~Zhang, P.~Myler, A.~Kelley, E.~Analau, S.~N.
  Hewitt, A.~J. Napuli, P.~Kuhn, R.~D. Ruth, and L.~J. Stewart.
\newblock {\em Physical Review Accelerators and Beams}, 11:91--100, 2010.

\bibitem{betal2009}
M.~Bech, O.~Bunk, C.~David, R.~Ruth, J.~Rifkin, R.~Loewen, R.~Feidenhans'l, and
  F.~Pfeiffer.
\newblock {\em Journal of Synchrotron Radiation}, 16(1):43--47, Jan 2009.

\bibitem{setal2012}
S.~Schleede, F.~G. Meinel, M.~Bech, J.~Herzen, K.~Achterhold, G.~Potdevin,
  A.~Malecki, S.~Adam-Neumair, S.~F. Thieme, F.~Bamberg, Ko. Nikolaou,
  A.~Bohla, A.~{\"O}. Yildirim, R.~Loewen, M.~Gifford, R.~Ruth, O.~Eickelberg,
  M.~Reiser, and F.~Pfeiffer.
\newblock {\em Proceedings of the National Academy of Sciences},
  109(44):17880--17885, 2012.

\bibitem{aetal2013}
K.~Achterhold, M.~Bech, S.~Schleede, G.~Potdevin, R.~Ruth, R.~Loewen, and
  Pfeiffer F.
\newblock {\em Scientific Reports}, 3:1313, 2013.

\bibitem{k2004}
G.~A. Krafft.
\newblock {\em Physical Review Letters}, 92:204802, May 2004.

\bibitem{tdhk2014}
B.~Terzi\'c, K.~Deitrick, A.~S. Hofler, and G.~A. Krafft.
\newblock {\em Physical Review Letters}, 112:074801, Feb 2014.

\bibitem{tetal2019}
B.~Terzi\'c, A.~Brown, I.~Drebot, T.~Hagerman, E.~Johnson, G.~Krafft, V.~Petrillo and M.~Ruijter.
\newblock {\em European Physical Letters}, accepted, 2019.

\bibitem{hsk2010}
T.~Heinzl, D.~Seipt and B.~K\"ampfer.
\newblock {\em Physical Review A}, 81:022125, Feb 2010.

\bibitem{gsu2013}
I.~Ghebregziabher, B.~A.~Shadwick and D.~Umstadter.
\newblock {\em Phys. Rev. ST Accel. Beams}, 16:030705, Feb 2013.

\bibitem{trk2016}
B.~Terzi\'c, C.~Reeves and G.~A.~Krafft.
\newblock {\em Phys. Rev. Accel. Beams}, 19:044403, Apr 2016.

\bibitem{metal2018}
C.~Maroli, V.~Petrillo, I.~Drebot, L.~Serafini, B.~Terzi\'c, and G.~A. Krafft.
\newblock {\em Journal of Applied Physics}, 124(6):063105, 2018.

\bibitem{ketal2016}
G.~A.~Krafft, E.~Johnson, K.~Deitrick, B.~Terzi\'c, R.~Kelmar, T.~Hodges, W.~Melnitchouk
and J.~Delayen.
\newblock {\em Phys. Rev. Accel. Beams}, 19:121302, Dec 2016.

\bibitem{ketal2018}
J.~M. Kr\"amer, A.~Jochmann, M.~Budde, M.~Bussmann, J.~P. Couperus, T.~E.
  Cowan, A.~Debus, A.~K\"ohler, M.~Kuntzsch, A.~Laso~Garc\'ia, U.~Lehnert,
  P.~Michel, R.~Pausch, O.~Zarini, U.~Schramm, and A.~Irman.
\newblock {\em Scientific Reports}, 8, 2018.

\bibitem{setal2009}
C.~Sun, J.~Li, G.~Rusev, A.~P.~Tonchev and Y.~K.~Wu.
\newblock {\em Phys. Rev. ST Accel. Beams}, 12:062801, Jun 2009.

\bibitem{cetal2017}
C.~Curatolo, I.~Drebot, V.~Petrilo and L.~Serafini.
\newblock {\em Phys. Rev. Accel. Beams}, 20:080701, Aug 2017.

\bibitem{retal2018}
N.~Ranjan, B.~Terzi\'c, G.~A.~Krafft, V.~Petrillo, I.~Drebot and L.~Serafini.
\newblock {\em Phys. Rev. Accel. Beams}, 21:030701, 2018.

\bibitem{f1991}
M.~Furman.
\newblock {In {\em Proceedings of the 1991 Particle Accelerator Conference}}, 422, 1991.

\end{thebibliography}
\end{document}